\documentclass[11pt,a4paper]{article}
\pdfoutput=1 

\usepackage{jinstpub} 
\usepackage{indentfirst}
\usepackage{graphicx}
\usepackage{gensymb}

\title{On the Electrode Configurations in a Large Single Phase Liquid Xenon Detector
	for Dark Matter Searches
}

\author[1]{P. Juyal,\note{Corresponding author.}}
\author{K. L. Giboni,}
\author{X. Ji}
\author{and J. Liu}

\affiliation{School of Physics and Astronomy, Shanghai Jiao Tong University,\\
MOE Key Laboratory for Particle Astrophysics and Cosmology, Shanghai Key Laboratory for Particle Physics and Cosmology, Shanghai 200240, China}

\emailAdd{pratibhajuyal@sjtu.edu.cn}

\abstract
{In the near future there will be the request for very large liquid Xenon (LXe) detectors for Dark Matter (DM) searches in the 50-ton range. To avoid an impractically long, single drift space of a dual-phase detector, it seems beneficial to use the single-phase technique. Since electrons then can drift in any direction, we can segment the homogeneous medium and thus avoid an excessive maximum drift path of order 4 m. The shorter detector length has several benefits, e.g. requiring a lower cathode voltage for the same drift field. We can easily split the TPC into two regions with the cathode in the center and two anodes at the top and bottom. One also can use multiple TPCs stacked on top of each other in the same liquid volume to reduce the maximum drift length even further. 

A further division of the drift space by installing an additional anode in the center would require S2 photons to traverse the liquid for several times the Rayleigh scattering length in LXe, which is only 30 - 40 cm. This seems to be excessive for good x - y localization. We therefore suggest a geometry of two independent TPCs with two drift spaces each. 

Despite earlier publications concerns persisted about the effect of shadowing. A detailed FEM model of the anode regions shows that with an aligned wire arrangement the drifting electrons impinge sideways on the anode in a narrow angular range of width 15$^{\circ}$ - 20$^{\circ}$. Most S2 photons are emitted in full view of the close-by PMT array. About 37\% of the S2 photons are shadowed by the anode wire out of which 30\% will be reflected back again on the gold plating of the wires. Thus we can observe 74\% of the total S2 light. Compared to a dual-phase detector, however, we do not suffer from the extraction efficiency, sometimes reported as low as 50\%.

}


\begin{document}
\maketitle
\flushbottom

\section{Introduction}
Since several years the search for dark Matter (DM) in the form of Weakly Interacting Massive Particles (WIMPs) is one of the focal points of physics research. The strong interest in DM spawned the development of ever larger LXe experiments, with the steady progress well reflected in the size of the deployed detectors. Of course, the positive identification of a signal depends not only on the target mass, but also on background rejection capability. And we saw an equally impressive progress in background rejection and control. From the modest 5 - 10 kg of early detectors in 2005, we have seen several generations every time increasing the active mass by about a factor 10. Within a few months we shall see the first results in the 5 - 10-ton range, namely the 4-ton PandaX IV~\cite{1}, the 8-ton XENON nT~\cite{2}, and the 10 ton LZ~\cite{3}. Despite the more than decade long concerted effort discovery of WIMPs eluded all attempts.

With the enhanced sensitivity at low energies, slowly the irreducible neutrino background starts to show up. Observing this background and identifying its structure we might be able to subtract the sources and gain more information on the elusive WIMPs. And of course going lower in the cross section might reveal a wealth of new physics beyond the Standard Model. This physics would be interesting in its own right, and new detectors should be geared to detect all the new phenomena additional to a possible WIMP discovery.  

WIMP detection is not a threshold effect. If no positive signal is identified in the next generation of detectors we shall request a large increase in target mass for a statistically significant improvement of the sensitivity. On the other hand, if there should be an observation, we have to further enhance the signal to study the physics of the find. Thus, in either case we shall require a future detector with at least 10 times the target mass. In this contribution we discuss a novel design concept of a 60-ton detector.

\section{Single-Phase Detector}

\subsection{Measurement Principle}
Xenon is a good medium for rare event searches at low energies. As a liquefied noble gas it has a density of about 3 g/cm${^3}$ and a high Z. This keeps a detector compact and easier to design and operate. Xenon has no long lived isotopes resulting in a so-called `self-shielding' ability, i.e. outer layers of xenon shield the inner part of the detector from ${\gamma}$ and X- ray background. Liquid xenon (LXe) can be used to detect the scintillation light or the liberated ionization electrons, and the combination of the two methods proved very successful for background discrimination. Finally, like all liquid detectors it is homogeneous and scalable in size over a large range. The properties of LXe detectors were reviewed in ~\cite{4} ~\cite{5}.

All the large LXe DM detectors are time projection chambers (TPC) measuring the ionization charge (S2) and the scintillation light (S1), with the exception of XMASS ~\cite{6}, which was a pure scintillator. Because the produced charges in the energy range of interest are too small to be observed with a charge sensitive amplifier, all these detectors used the so-called dual-phase principle. It was originally proposed by Dolgoshein ~\cite{7} and was first suggested in ~\cite{8} for WIMP searches in 1989. The drifting electrons are extracted from the liquid to be detected in the gas phase by their electro-luminescence with the same photo multiplier tubes (PMT) as the scintillation light. 

Early on the existence of electro-luminescence in the liquid phase was demonstrated by the Waseda group~\cite{9}~\cite{10}. Most relevant in the context of this study is that the drifting electrons do not have to cross the liquid level, i.e. their direction of drift is not limited to upward. They might as well drift down or sideways. Thus, we can divide the active volume into multiple drift regions or independent TPCs stacked on top of each other. Specifically, we aim at a design with 4 drift spaces. Compared with the equivalent detector geometry of a conventional dual-phase detector, we can lower the cathode HV by a factor 4 for the same drift field. The maximum drift time is also reduced by the same factor, and so is the attachment of drifting electrons to electro -negative impurities.

We propose to use the same PMTs ~\cite{11} as in present detectors, since they were specifically designed for a LXe environment with special attention to low internal radioactivity. We note, however, that different photo sensors might offer the big advantage of shorter length, e.g. silicon photomultipliers (SiPMs), sometimes also called multi-pixel photon counter (MPPCs). But additional development might be required, especially to fulfill the low radioactivity requirement. For SiPMs in particular, also the high dark count rate at LXe temperature has to be reduced to avoid raising the detection threshold. As now customary, the side walls of the detector are covered with polytetrafluoroethylene (PTFE or Teflon) reflectors~\cite{12}. We thus forego the use of pulse shape discrimination (PSD) which appears promising in simulation~\cite{13}, but would require to avoid all reflections and cover the side walls with very fast PMTs. The impact of PSD in LXe as a background discriminator is not experimentally established. This and other benefits of a 4${\pi}$ light detection system might be interesting for future upgrades.

\subsection{Assumed Geometry}
\begin{figure}[!hbp]
	\centering
	\includegraphics[width=1\textwidth]{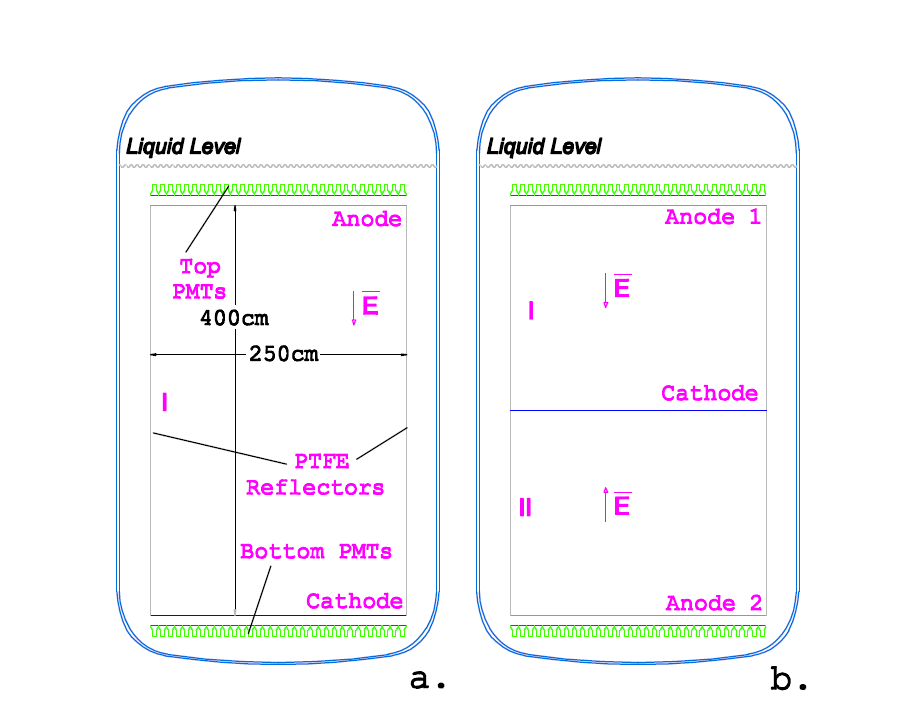}\\
	\caption{Schematic designs of the (a) Dual-phase, (b) Single-phase with two drift spaces}
	\label{fig 1}
\end{figure}

\setlength{\parindent}{5ex}The geometry of the assumed detector provides a cylindrical active volume of 2.5 m diameter and 4 m height, corresponding to roughly 60 tons of LXe. A conventional dual-phase detector with these dimensions is shown in Fig.~\ref{fig 1}a for comparison. The active volume is delineated by the cathode on negative HV below and the anode structure a few mm above the liquid level. The active volume is viewed by the top and bottom PMT arrays with total 1914 PMTs. A conventional field cage surrounds the active volume providing the homogeneous electric field. The cathode and the field cage determine the electron drift, but have no direct influence on S2 production which will be our main theme. 

With the single-phase technique, we can easily achieve two drift spaces by removing the cathode to the center and introducing a second anode structure in front of the bottom PMTs, as shown in Fig.~\ref{fig 1}b. The liquid level, so critical in its position in a dual-phase, is somewhere above the top PMTs. The detector is now entirely symmetric. Also precise leveling of the detector is no longer required.

In section~\ref{Multiple Drift Spaces}, we shall introduce the division of the active volume into 4 drift spaces.

\section{Anode Configuration}
\label{Anode Configuration}

\begin{figure}[!ht]
	\centering
	\includegraphics[width=0.8\linewidth]{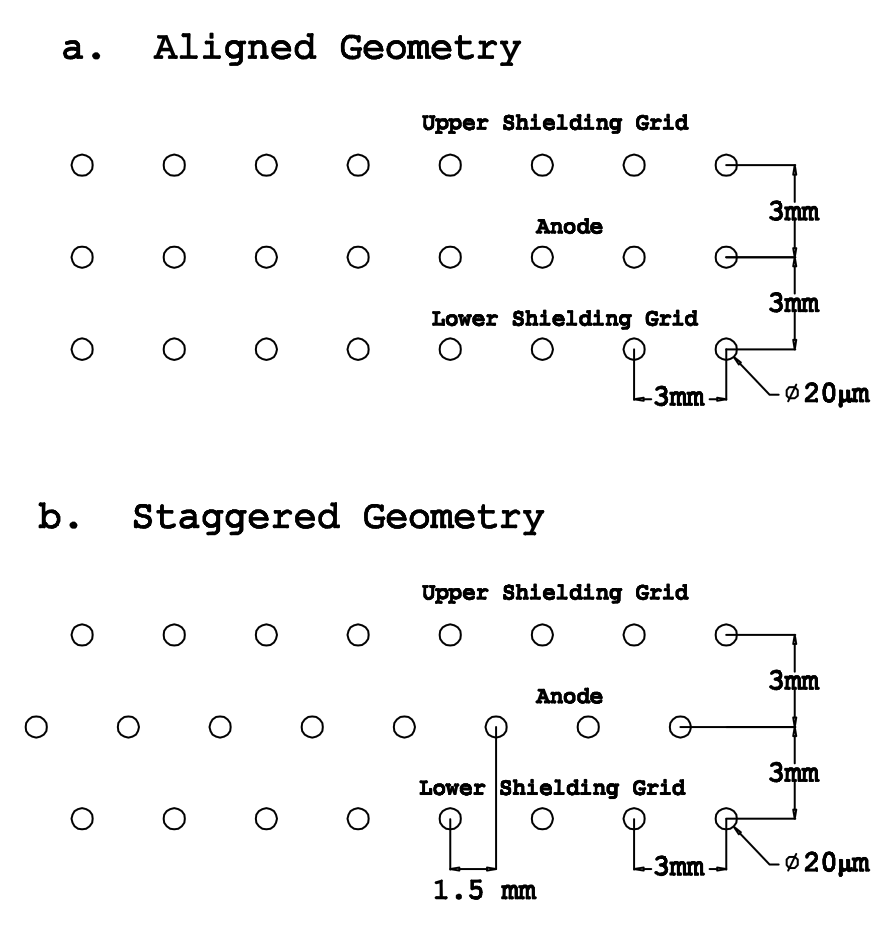}\\
	\caption{(a) Aligned and (b) Staggered arrangement for the wires of the anode and the two shielding grids. }
	\label{fig 2}
\end{figure}

Single-phase LXe detectors were recently reevaluated in two tests by the groups from SJTU~\cite{14} and from the Columbia Astrophysics Lab~\cite{15} (CAL). The geometries were quite similar except for the anode wire configuration. One employed a single anode wire located in between the position of the adjacent grid wires, whereas the other had a wire grid as anode, aligned with the grid wires. Although the results at first appeared to disagree, they are consistent~\cite{16} with each other and with the earlier results from the Waseda group. The two wire arrangements are schematically presented in Fig.~\ref{fig 2}. We shall look at their properties with detailed finite element modeling. 

All our simulations assume anode electrodes of stretched gold-plated tungsten wires, a standard element of multi wire proportional- and drift- chambers. The wire diameter is 20 $\mu$m, and the spacing between wires and between wire planes is 3 mm. The shielding grids sandwiching the anode can also be made with thicker wires, e.g.  50 - 100 $\mu$m without any change. They are solely used for field shaping and not to achieve a position independent detection like the well-known Frisch ~\cite{17} grids in gridded ionization chambers. 

\subsection{Aligned Anode}
\label{Aligned Anode}
\setlength{\parindent}{1cm}
\indent

We first study the geometry of Fig.~\ref{fig 1}b. Here the distance between anodes and PMT arrays is only a few cm, much shorter than the Rayleigh scattering length. We note the detector is now fully symmetric, and all results will hold for the top and bottom anode. In our previous study ~\cite{16}, we qualitatively concluded that the observed S2 with the aligned geometry of Fig.~\ref{fig 2}a will better resemble the response of a conventional dual-phase detector. In this geometry the anode wires are parallel and located in between two wires of the shielding grids above and below. The anode potential is chosen such that the field strength at the wire surface just reaches the threshold for electron multiplication. With the grid wires on ground the anode potential can be determined from the 1/r dependence to 4140 V. The drift field is always much lower than the field in the anode gap and the grids are totally transparent~\cite{18}. The drifting electrons are deviated from their straight line path in the drift region and compressed into the opening between grid wires. They follow the field lines and will hit the anode wires preferentially from the side. To quantify the effect, we modeled the field in great detail, ensuring that the meshing of the finite element model accounted for the wire diameter of only 20 $\mu$m. 

\begin{figure}[!ht]
	\centering
	\includegraphics[width=1\textwidth]{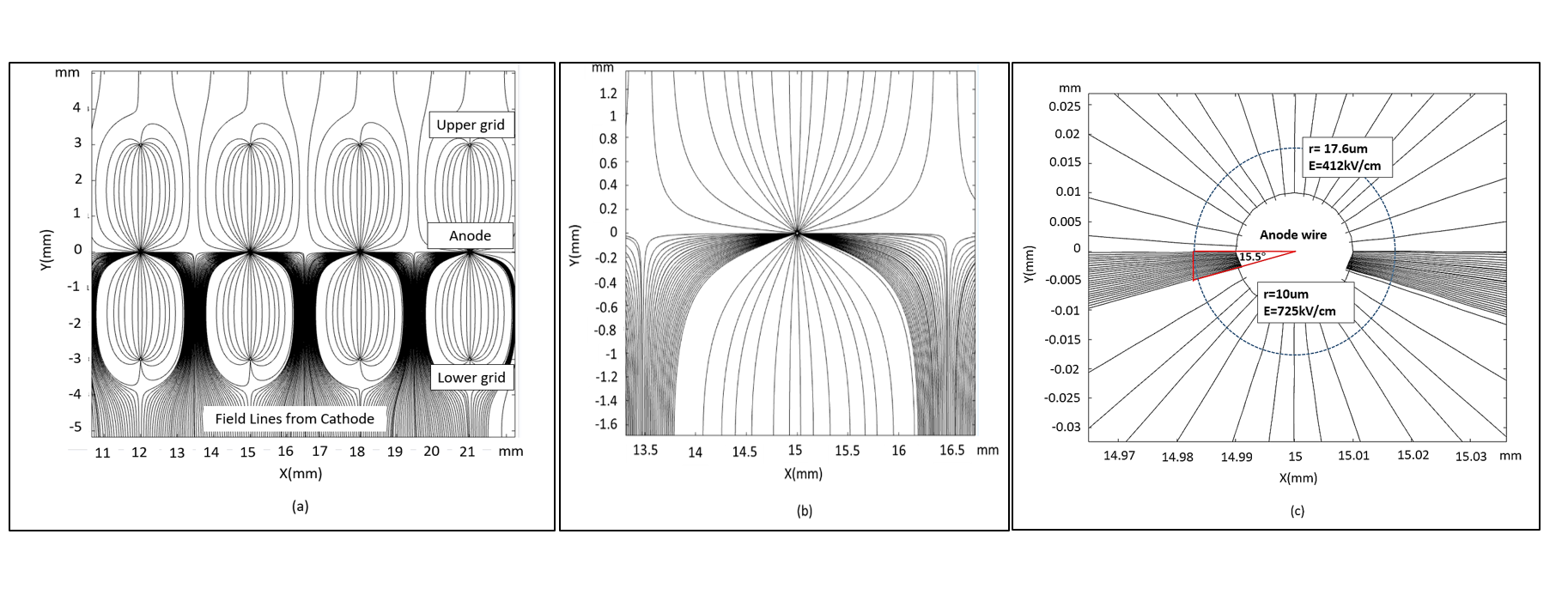}\\
	\caption{a.) Field line distribution in the anode region with aligned wires of the detector. The wire diameter is 20  $\mu$m. The spacing between wires is 3 mm. The dense field lines originate on the cathode of the detector and are focused by the grid wires.  b.) Zoom in around a single wire. The lines from the cathode are compressed to a very small region and hit the anode wire from both sides.  c.) The region in proximity of the wire. The threshold for electro-luminescence is at 17.6 $\mu$m (dashed circle) from the wire center. Only electrons within the red triangle contribute to the S2 signal.}
	\label{fig 3}
\end{figure}

The overall field distribution in the anode region is displayed in Fig.~\ref{fig 3}a. The applied potentials were 4140 V on the anode with the shielding grids on ground potential. The cathode was -800 V. Only the region of the anode structure with the two shielding grids is shown. As expected the lower grid focuses the drifting electrons which continue their upwards path until they are forced sideways by the field from the upper grid wires. In the plot we intensified the number of displayed field lines from the cathode to better identify their paths in the zoom of Fig.~\ref{fig 3}b. The lines from the cathode appear like a black band. The low density lines in Fig.~\ref{fig 3}b originate at the lower grid wire, and no drift electrons follow their path.

In a yet larger magnification Fig.~\ref{fig 3}c shows the close vicinity of the anode wire. Electro-luminescence will start at a threshold of 412 kV/cm, indicated by the dashed circle. Electrons from the drift space below will approach the wire in a narrow angular range of $\theta$ = 15.5$^{\circ}$, symmetrically from both sides. Only electrons within this region indicated in red and within the dashed circle at 17.6 $\mu$m can contribute to a S2 signal. The radial boundary of this region is determined by the wire diameter and the anode potential. Choosing the anode potential such that the multiplication threshold is crossed at the wire, the radius of the region with electro-luminescence is given by the 1/r field distribution. Proportional scintillation starts at a field (412 kV/cm) very high compared to the average field in the gap of order 10 kV/cm. This means the radius is quite insensitive to the exact wire locations. The angular range, on the other hand, is a consequence of the focusing by the shield wires. By choosing a typical drift field of 1 kV/cm the grid is completely transparent, however the focusing depends on the ratio the electrons see before and after the grid. Due to the length of the drift region 1 kV/cm is already on the high side of any typical design. This means the focusing in a real detector will normally even stronger.

\begin{figure}[!ht]
	\centering
	\includegraphics[width=0.8\textwidth]{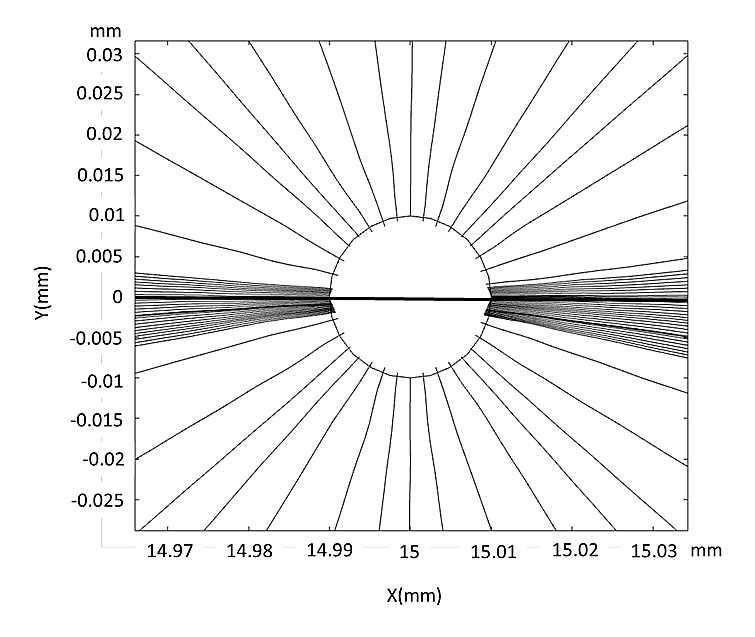}\\
	\caption{Field line distribution as in Fig.~\ref{fig 3}c, but with asymmetric grid potentials. The bottom grid is on ground potential and the upper grid at +300V. The thick line shows the horizontal. Compared to Fig.~\ref{fig 3}c the angular range of incidence is now turned upward.}
	\label{fig 4}
\end{figure}

The angle of incidence of the electrons on the wire is always below the horizontal. To enhance the signal on the top PMT one can turn the range of hits on the wire by changing the bias of the top grid, leaving all other potentials unchanged. Fig.~\ref{fig 4} demonstrates the effect of asymmetric potentials. The top grid is now on +300V while the lower grid is still on ground. The angular range is turned slightly upward. The grid voltages can thus be varied to fine tune for an optimal performance. The short distance of 7.6 $\mu$m can be passed by the electrons in a very short time. We therefore expect very short S2 light pulses.

\subsection{Staggered Anodes}
\label{Staggered Anodes}
\begin{figure}[!ht]
	\centering
	\includegraphics[width=1\textwidth]{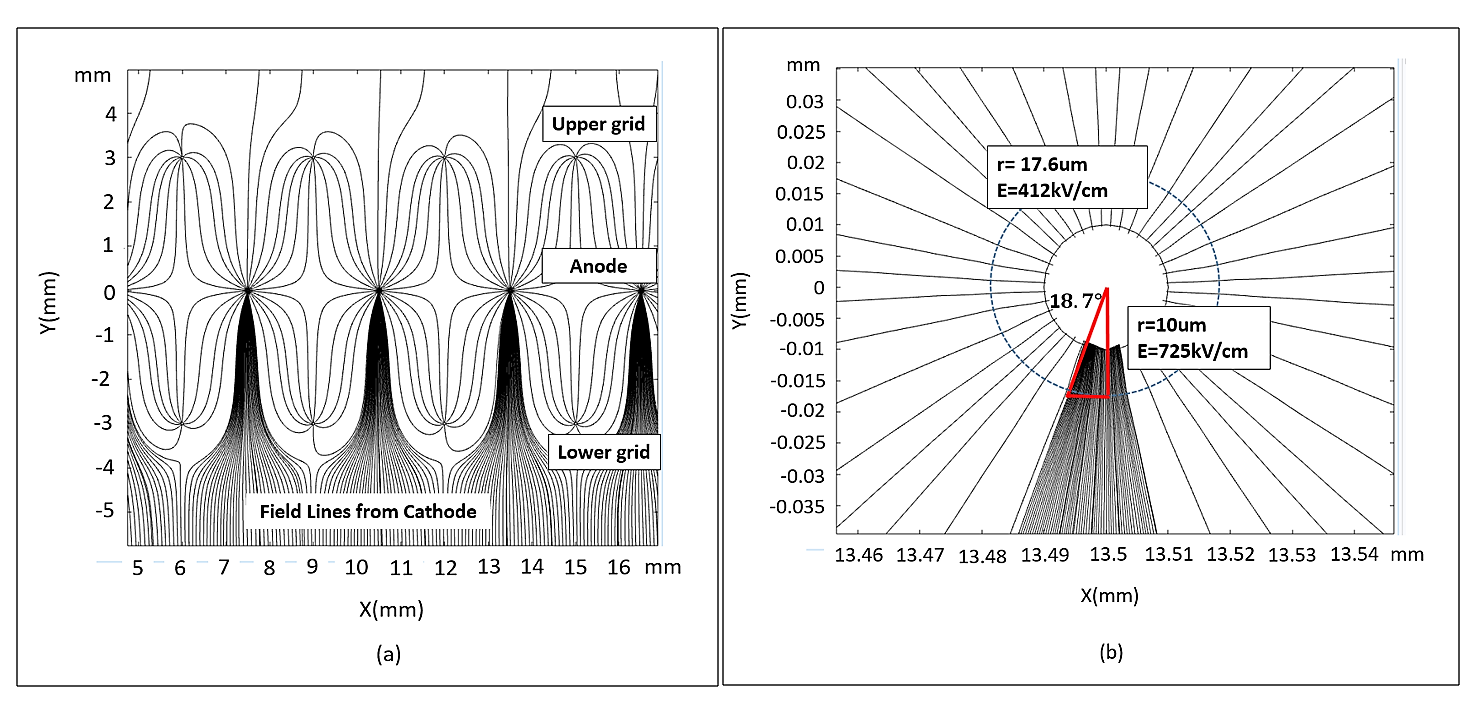}\\
	\caption{a.) Field line distribution in the anode region with staggered wires as in Fig.~\ref{fig 2}b. The wire diameter and the pitch are the same as before. The field lines from the cathode are again focused by the grid wires and impinge on the anode from below. The region of electro-luminescence is indicated by the dashed circle. b.) The region in proximity of the anode wire. The two distinct ranges at opposite sides of the anode wire are now adjacent to each other.}
	\label{fig 5}
\end{figure}

A second anode configuration of interest has staggered wires as shown in Fig.~\ref{fig 2}b. Naturally, the shift in wire positions changes the field distribution and the electron trajectories. Fig.~\ref{fig 5}a shows the simulation of the fields analogous to the previous section. The field lines originating on the cathode are again bent around the grid wire. Now they are focused onto the anode wire from below. In Fig.~\ref{fig 5}b we see the vicinity of the anode wire in a zoom to determine the angle of incidence. The angular range has a slightly larger width than before with 18.7$^{\circ}$, but the two ranges from opposites sites are now adjacent. Drifting electrons in an event are only in one of the two angular ranges, not in both. To generate S2 photons the field has again to be above the threshold of 412 kV/cm indicated by the dashed circle. The potential on the electrodes are the same as in the previous case, grids on ground and anode at 4140 V. Increasing the anode potential would result in a stronger focusing action, and the angular range might be slightly smaller.

\subsection{Shadowing and S2 Efficiency}
\label{Effects of Shadowing and S2 Efficiency}
 In close vicinity of the anode wire the drift electrons isotropically produce electroluminescence proportional to the local electric field strength. We have chosen the anode potential to always stay below the threshold for electron avalanches. This will avoid additional fluctuations from the multiplication process. The length of the electron path is only 7.6 $\mu$m. At such short distances the anode wire will always cast a shadow and reduce the S2 signal. In the data analysis the S2 light is used for two purposes, total charge measurement and position determination in x-y.

\subsubsection{Total Charge Measurement}
We simulated drift electrons according to a geometry of Fig.~\ref{fig 2}a. The photons were integrated over their path to include the production mechanism obtained from the field distribution of Fig.~\ref{fig 3}. The average number of photons is about 20 per electron. About 37\% of all photons are impinging on the wire. But some of these photons are reflected on its gold coating. These also contribute to the total charge measurement with the S2 signal. The reflectivity  ~\cite{19}~\cite{20}  of gold at the xenon wavelength is roughly 30\%. Thus, 74\% of the S2 light is available for detection despite the shadowing effects.
 
The group from CAL ~\cite{15} increased the anode potential to achieve an S2 gain comparable to a dual-phase detector of about 200 - 300 photons/electron. They included a factor 14 from electron multiplication in their signal formation. On the other hand, they used very thin wires (10 $\mu$m), reducing the effective path length and thus the electro-luminescence by 50\%. 

Increasing the wire diameter, e.g. 50 $\mu$m, would increase the gain, but a sturdier electrode frame might be required to support the mechanical forces of the stretched wires. This might introduce additional radioactive background. Alternatively, the anode potential can be raised beyond the electron multiplication threshold. This does increase fluctuations from electron avalanches. But at very low energies, i.e. our main region of interest, the added fluctuations are outweighed by the improved counting statistics. The best solution has to be determined in the final optimization of a design.

We also can raise the anode potential. We had chosen a reduced field strength to avoid fluctuations from electron avalanches. But at very low energies, i.e. our main region of interest, the added fluctuations from small gain enhancements are outweighed by the improved counting statistics. 

The overall light yield of S2 depends on many design parameters such as the quantum efficiency and the noise performance of the PMTs. In the large arrays the light is shared by many PMTs, and at low energies most of them will only detect a single photo electron. We also have to consider that in a dual-phase scheme the number of electrons is reduced by the extraction efficiency. Since the anode is in the gas phase, spurious breakdowns enforce a reduction of the anode potential for stable, long time operation. Practically the extraction is often less efficient than the calculated design parameter to improve the long term reliability of the measurement. Sometimes the actual efficiency is reported ~\cite{21} to be of order 50\%.

The total light collection efficiency of a single-phase detector can be comparable to a dual-phase. It depends on the final optimization of several detector parameters, like for example the photo cathode coverage of the top array. In order to have a fully symmetric geometry in our design we have chosen a tightly- packed hexagonal array, identical to the bottom array.

\subsubsection{X-Y Position Determination}
The second use of the S2 light is to determine the x-y location of any event. We note that the z coordinate is determined from the timing of the S2 signal relative to S1, which is used as t$_{0}$. The timing information also shows if there was only a single interaction or more than one ionizing event, as for example in Compton scattering. For any event of interest, we therefore are certain that it occurred in a single point in the detector.

Like in a dual-phase detector the S2 light forms a narrow cluster in the PMT array which is at a short distance of only 5 - 10 cm. It now becomes essential that the electrons hit the anode wire from the side. The shadow of the wire is in horizontal direction, i.e. it falls into a region with no PMTs or PTFE reflectors. With the asymmetric grid voltages of Fig.~\ref{fig 4} we even can turn the range of incidence above the horizontal to ensure that all the upward going photons hit the top PMT array. Naturally, the same is true for downward going photons at the bottom of the detector. Reflections from the gold surface of the wires will also contribute to the cluster in the top, or bottom, array. 

Since sufficient S2 light is contained in the cluster for x-y determination, in a dual-phase detector the event location is normally derived only from the top array. Our single-phase geometry of Fig.~\ref{fig 1}b provides a comparable hit cluster in the close-by PMT array. It will not be necessary to include the light distribution of the opposite PMT array in the x - y localization.

It is sometimes reported that the x-y position of an event can also be determined from the opposite PMT array alone, i.e. the bottom array in a dual-phase. These observations are obtained in much smaller detectors. For a very large detector it is not obvious that this is still valid. The S2 photons on their long path through the detector are subject to Rayleigh scattering and possibly multiple reflections on the PTFE walls. Since the Rayleigh scattering length in LXe is of order 30 - 40 cm, the original position information of these photons might be entirely lost on a 2 m path length. How much of the original information is retained and if it can contribute to an accurate x-y localization has to be evaluated by simulation and verified by experimental tests.

\subsection{Active Shield Region}
\label{Active Shield Region}
Present large size dual-phase detectors already use the feature of self-shielding in LXe. The field cage of a detector has to be surrounded at the sides by several centimeters of `dead' xenon to stand off the HV from the field shaping electrodes and the cathode. XENON100 ~\cite{22} was the first detector to instrument the liquid xenon around the active volume with dedicated PMTs. This outer layer is thus turned it into an active shield against $\gamma$- and X- ray backgrounds, e.g. from the radioactivity of the detector vessel.

The strongest background in a LXe detector stems from the low activity PMTs itself. In a dual-phase detector, it is not possible to shield this background since the anode has to be in low density gas. In the single-phase geometry according to Fig.~\ref{fig 1}b all the volume in front of the PMTs is filled with dense liquid. Photons from the PMTs are attenuated as well. We cannot avoid to observe S1 signals from background interacting in the shield region, but no S2 will be produced if there is no electric field.

The region in front of the PMTs is field free if they are powered on the anode with positive HV, i.e. the photo cathode and the case of the PMT are on ground. In case of negative PMT HV we easily can shift the potentials of the anode and the shielding grids. We bias the shielding grids with a potential more negative than the PMTs. Of course, the anode potential has also to be reduced to keep the same field distributions. The detector geometry is entirely symmetric with respect to top and bottom. Furthermore the bottom PMT array is no more facing the HV of the cathode, i.e. the customary grid in front of the bottom PMTs is no more required.

\section{Multiple Drift Spaces}
\label{Multiple Drift Spaces}
Selecting a single-phase geometry alleviates the design of the mechanical support, since leveling and control of the liquid level are superfluous. All HV connections are now within the LXe, which is an excellent insulator. Thus, we expect a much better stability during long time operation. But it also grants much more freedom in the design. The construction of a multi-ten ton detector is now similar to the arrangement of several modules without increasing the number of read out channels. We mentioned that dividing the active volume will reduce the cathode HV, the diffusion of the drift electrons, and the attachment to electro-negative impurities linearly with the number of drift spaces. 
We studied a detector according to Fig.~\ref{fig 1}b with two drift regions, each 2 m long. Now we shall strive for another factor 2 by forming 4 drift spaces.

\subsection{Single TPC}
\begin{figure}[!ht]
	\centering
	\includegraphics[width=1\textwidth]{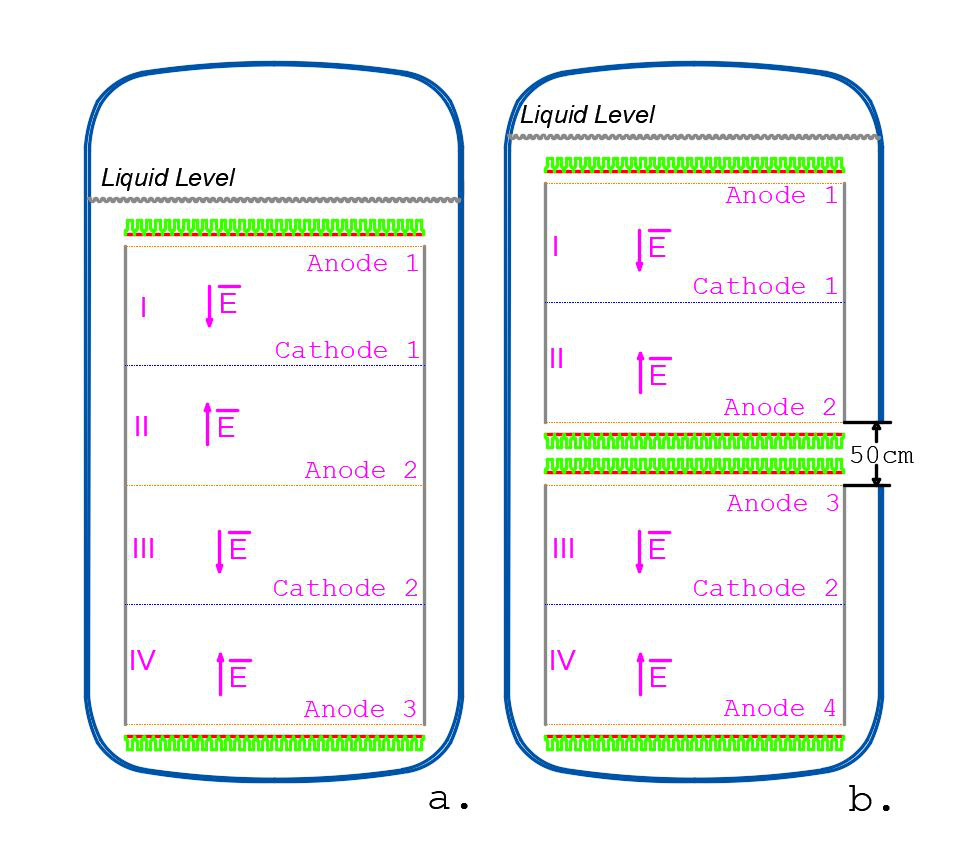}\\
	\caption{Schematic comparison of detectors with 4 drift spaces (a) Single TPC with additional anode in the center, (b) Twin TPC with two identical TPCs on top of each other.}
	\label{fig 6}
\end{figure}

A simple way of forming 4 drift spaces in our detector is to introduce an additional anode structure in the center and provide 2 more cathodes. Such a design is shown in Fig.~\ref{fig 6}a. We need only two PMT arrays, and the number of feedthroughs, cables, and electronic read out channels remains unchanged. This design has a major drawback. We can measure the z - position of an event with very good accuracy, but we have an ambiguity as to the drift space. The outer drift spaces (I and IV in Fig.~\ref{fig 6}a) can be identified by the narrow cluster in the hit pattern. For the other two spaces all the S2 light is produced at the central anode. An event in space II would be indistinguishable from one at the same distance in space III.

To resolve this ambiguity, we can change the center anode to a staggered wire geometry shown in Fig.~\ref{fig 2}b. In section~\ref{Staggered Anodes} we learned that the angular range of the electron impact is nearly the same as for aligned anodes, but the electrons approach the anode vertically from only one direction. The shadow of the wire will cover most of the PMT array in this direction, but it will not affect the opposite one. If we include reflections on the wire we expect about the same light yield, but concentrated on the opposite PMT array. The strong asymmetry in light intensity will identify the drift space, II or III, which contained the event. Within a drift space the z - position is of course still measured with high accuracy by the drift time.

Although we can measure the S2 amplitude and z - position as usual, it might be challenging to achieve a good x-y which is required for the radial fiducial cut. Prior studies with small detectors showed that the x-y can be extracted from the hit pattern at some distance. But in our case this distance has grown to more than 2 m, far beyond the Rayleigh scattering length in LXe of 30 - 40 cm. Additional multiple reflections on the PTFE walls might wash out the remaining position information.

Although we have an attractive geometry it might only be used after additional simulation studies and experimental verification that the x-y localization is still sufficient.

\subsection{Twin TPC}

We still can form 4 drift spaces in a less elegant but risk free way with a Twin TPC detector according to Fig.~\ref{fig 6}b. Each of the two TPCs is independent of the other with its own PMT arrays. They have only half the height, and are stacked on top of each other. We realize that we need twice the amount of PMTs and read out channels. And we have to accept a gap between the detectors for the arrays. With the presently favored PMTs we expect this dead space to amount to 50 cm, mostly determined by the length of the PMTs.

In the future we hope that one can reduce the dead space to less than 10 cm, when novel photon sensors will become available. There are two very promising developments:

a. A 2" square tube in metal-channel design ~\cite{23} with an overall length of 25 mm is already available with reduced internal  radioactivity. But the value ~\cite{24} is still higher than the commonly used PMT. A reduction might be possible in the future. 

b. We also would expect that SiPMs, will come of age for DM detectors before a 60 ton can be realized. 

Both solutions would also reduce the buoyancy of the photon sensor arrays. At present with about 300 g per PMT it requires a sturdy TPC frame with the implied increased background.

\section{Conclusion and Outlook}
Adopting the single-phase LXe scheme has many advantages for the design, operation, and stability of the detector. Removing the constraint that the drifting electrons must pass the liquid level has many consequences. The most important difference is that electrons can drift in any direction, i.e. we can configure multiple drift spaces.

From detailed modeling of the close vicinity of the anode wire we conclude that the drifting electrons approach the anode wire sideways in a narrow angular range of typically 15$^{\circ}$  width. The photons hitting the wire are not all lost. The gold coating of the conventional tungsten wires reflects about 30\%. We estimate that the amount of observable photons is 74 \%, but we do not suffer from the electron extraction efficiency of a dual-phase detector. During operation this efficiency might be much lower than the design parameter to avoid spurious HV breakdowns in the gas around the anode. An important advantage of the single-phase method is that the space between active volume and top PMTs is filled with dense liquid. Thus the background from the residual radioactivity of the top array is also attenuated by the self-shielding.

We propose a geometry with 4 drift spaces resulting in a reduction of the cathode HV by a factor 4. At the same drift field the attachment to electro-negative impurities of the drifting electrons is approximately reduced by the same factor, while electron diffusion is reduced by a factor of 2. For the anodes on top and bottom we recommend to align the anode wires with the shielding grids. An elegant realization of the TPC would have an additional anode structure in the center of the detector. An ambiguity in the drift space occupied by the event can be resolved by using a staggered anode geometry. But it is not obvious that one can achieve an adequate x-y localization in the center drift spaces for the required fiducial cuts. The path length for all S2 photons from the center anode would be in excess of 2 m, which does not well compare with a Rayleigh scattering length of about 30 cm in LXe. The hit position in the PMT array would be further randomized by multiple reflections on the PTFE walls.

We thus prefer a geometry with two independent TPCs of half the length mounted on top of each other. In between the TPCs two additional photon sensor arrays would eliminate the x-y positioning challenge. The dead space between the two TPCs would be about 50 cm. However, novel photon sensors such as SiPMs might reduce this to less than 10 cm.

In conclusion we would like to point out that our proposed detector is neither accepted or under serious consideration in any of the groups active in the field. More simulations and experimental verification will be required to transfer the proposed geometry from small laboratory tests to the scale intended.

\section{Acknowledgment}
This project is supported in part by a grant from the Ministry of Science and Technology of China (No. 2016YFA0400301), and Office of Science and Technology, Shanghai Municipal Government (grant No. 18JC1410200).

\end{document}